\documentclass[prl,preprint,aps,superscriptaddress]{revtex4}
\usepackage{amsmath}
\usepackage{graphicx}
\usepackage{tabularx}
\usepackage{amssymb}
\usepackage{hyperref}
\usepackage[usenames]{color}
\definecolor{rltred}{rgb}{0.75,0,0}
\definecolor{rltgreen}{rgb}{0,0.5,0}
\hypersetup{colorlinks,linkcolor=rltred,citecolor=rltgreen}

\begin{document}


\title{Geometrical optimization approach to isomerization:
Models and limitations}

\affiliation{School of Chemistry (BK21), Seoul National University, Seoul 151-747, Republic of Korea}

\author{Bo Y. Chang}
\affiliation{School of Chemistry (BK21), Seoul National University, Seoul 151-747, Republic of Korea}

\author{Seokmin Shin}
\affiliation{School of Chemistry (BK21), Seoul National University, Seoul 151-747, Republic of Korea}

\author{Volker Engel}
\affiliation{Institut f\"ur Physikalische und Theoretische Chemie,
Universit\"at W\"urzburg, 97074 W\"urzburg, Germany}

\author{Ignacio R. Sola}
\affiliation{Departamento de Qu\'imica F\'isica I, Universidad Complutense, 28040 Madrid, Spain}
\email{isola@quim.ucm.es}

\begin{abstract}
We study laser-driven isomerization reactions through an excited
electronic state using the recently developed Geometrical Optimization 
procedure 
[J. Phys. Chem. Lett. 6, 1724 (2015)]. The goal is to analyze
whether an initial wave packet in the ground state, with optimized
amplitudes and phases, can be used to enhance the yield of the reaction
at faster rates, exploring how the geometrical restrictions induced by the 
symmetry of the system impose limitations in the optimization procedure.
As an example we model the isomerization in an oriented
2,2'-dimethyl biphenyl molecule with a simple quartic potential. 
Using long (picosecond) pulses we find that the 
isomerization can be achieved driven by a single pulse.
The phase of the initial superposition state does not affect the yield.
However, using short (femtosecond) pulses, one always needs
a pair of pulses to force the reaction.
High yields can only be obtained by optimizing
both the initial state, and the wave packet prepared in the excited state,
implying the well known pump-dump mechanism. 
\end{abstract}


\maketitle


\section{Introduction}

Geometrical Optimization (GeOp) has been recently proposed as a
method to engineer the initial state within a manifold of
allowed or accessable initial levels in order to maximize the yield of
population transfer to a single state or a set of states belonging
to another manifold of levels\cite{par1,par2}.
The method has led to the discovery of Parallel Transfer (PT)
as a mechanism of efficient electronic absorption with
strong and short ({\it e.g.} femtosecond) laser pulses\cite{par1,par3}. 
PT accelerates the desired transition using weaker fields 
and can be used to achieve selective excitation with pulses
with bandwidths much larger than the vibrational spacing.
It has been argued that there are geometrical (or structural)
aspects of the Hamiltonian underlying the success of the PT process\cite{par3}.
In this work we investigate their role in the control of
isomerization reactions, where the geometry of the problem
has a clear direct impact on the process under study.

Quantum control has been used with great success to significantly
enhance the yields of photodissociation reactions by precise
tailoring of laser pulses\cite{QC1,QC2,QC3,QC4}. 
Many different strategies have been
suggested contributing to knowledge on key aspects of the quantum
dynamics, particularly under strong fields\cite{SS1,SS2}. 
Experiments typically use pulse-shaping technologies\cite{shaping1,shaping2} 
and learning algorithms\cite{LA} have been used in a wide variety of 
systems\cite{PDexp1,PDexp2,PDexp3,PDexp4,PDexp5,PDexp6,PDexp7}.

Numerical results using N-level 
Hamiltonians under long pulses and wave-packet calculations in reduced 
($1$ or $2$)-dimensional models for short pulse dynamics have also shown 
great promise in the possibility of driving isomerization reactions
\cite{pipulse1,pipulse2,pipulse3,spulse1,spulse2,spulse3,spulse4,spulse5,chirp2,STIRAP1,STIRAP2,STIRAP3,STIRAP4}
and even distinguishing optical isomers or purifying a racemate
mixture\cite{pipulse4,chirp1,CPL1,CPL2,CPL3,CPL4,CPL5,BS1,BS2,BS3,BS4,BS5,VE}.
The most general models of population transfer were applied.
For instance, population inversion via $\pi$ pulses were used with 
sequences of IR pulses\cite{pipulse1,pipulse2,pipulse3,pipulse4} 
or a single short IR pulse, acting in a pump-dump mechanism\cite{spulse1,spulse2,spulse3,spulse4}. 
Despite using relatively simple models, much attention
was devoted to analyzing the robustness of the schemes to different
types of intra- or inter-molecular couplings\cite{robustness,spulse2}.
Adiabatic passage using chirped pulses\cite{chirp1,chirp2}, 
STIRAP\cite{gSTIRAP1} (Stimulated Raman adiabatic passage)
\cite{STIRAP1,STIRAP2,STIRAP3,STIRAP4} 
or even half-STIRAP\cite{chirp1}, were also used.

The same approaches were also applied to the more difficult case of
symmetric isomerization reactions, as in optical isomers. 
Here the polarization plays an important role\cite{CPL1,CPL2,CPL3,CPL4,CPL5}, 
but the linear components of the polarized field can work separately if
there operates a mechanism for breaking the parity of the system\cite{BS1,BS2,BS3,BS4,BS5} or
in the simpler scenario where the molecule is aligned with the 
electric field\cite{pipulse4,chirp1}.
Even recent approaches with strong fields, based on the role of Stark effects
\cite{strong1,strong2,strong3,strong4}
or the use of counterdiabatic pulses to accelerate adiabatic passage\cite{counter}
were proposed and tested.

Despite the many suggested control mechanisms,
there is still limited success in controlling isomerization
reactions in experiments.
The experiments work with strong fields and use the excited electronic 
state as an intermediate of the reaction\cite{exp1,exp2,exp3,exp4,exp5,exp6,exp7}.
The reason is not only due to technological limitations in the 
set of IR sources available (and the difficulties in modulating these
pulses), but it is also motivated by important physical reasons.
In the Franck-Condon region of the excited state, the initial 
wave function experiences a natural force towards the reaction coordinate, 
that can in principle lead to fast isomerization in the absence of an internal 
barrier.
In addition, because of intramolecular vibrational redistribution (IVR) in 
and conical intersections of excited states (or other non-adiabatic
couplings) one typically needs to move the population rapidly through
the transition state, which favors doing it in the absence of a barrier
in the excited state.
Using strong fields to drive the electronic absorption leads naturally to 
study the effect of vibrational motion (or vibrational coherence)
to enhance such absorption\cite{IRUV1,IRUV2,IRUV3,IRUV4,IRUV5,IRUV6,IRUV7,IRUV8,IRUV9}. 
Recent results in two-photon processes
(such as a pump-dump mechanism) have shown that the optimization of
the initial wave packet is less important when the pulses are time-delayed\cite{GeOp2photon}.
In this work we investigate its role in isomerization reactions, with
the wider goal of finding new mechanisms to enhance the yield 
and especially accelerate the rate of the reaction.
To that end we study the {\em cis-trans} isomerization in 
2,2'-dimethyl-biphenyl. In Sec.2 we explain the model and summarize
how we apply the GeOp algorithm to the two-photon process. In Sec.3
we show the numerical results and build a simple analytical model
to explain the main observations. Sec.4 is the conclusions.

%

\section{Models and Methods}

As a simple general model for isomerization reactions, we use quartic
symmetric one-dimensional potentials with parameters fitted using spectroscopic data to obtain
potentials for the ground ($V_g(x)$) and first excited electronic state ($V_e(x)$), respectively.
In this work we analyze the dynamics
only in the reaction coordinate.  
Then we obtain the Hamiltonian in the energy representation, applying
the Fourier Grid Hamiltonian (FGH)\cite{FGH}. The transition dipole matrix 
elements are evaluated assuming the Condon limit.
Since we want to stress some geometrical features of the solution,
we use as a representative example the torsion of the phenyl groups
in the isomerization of the 2,2'-dimethyl-biphenyl, which we
represent by a double well potential.

\subsection{Parameters of the isomerization in 2,2'-dimethyl-biphenyl}

Although to model torsional potentials one often uses polynomials of trigonometric
functions ($\cos(m\theta)$) of the torsional angle $\theta$, we, for simplicity,
model the dimethyl-biphenyl ground state using a quartic expression.
This is also justified because the control schemes that we use are properly
described in the energy representation, so that we are mainly interested
in having a model as general as possible that approximately reproduces the 
energetics of the isomerization reaction. In that regard we will use
scaled units.
Writing
\begin{equation}
V_g(x) = \frac{1}{2}\beta x^2 \left[ \frac{x^2}{2x_0^2} -1 \right]
\end{equation}
where $x$ is a generalized reaction coordinate, such that
$2x_0$ is the separation between the equilibrium configurations of
both isomers (approximately the torsional angle displacement between
the isomers, here $90.7^\circ$\cite{Grein}), 
and $\beta$ plays the role of the spring constant of the oscillator in
each equilibrium configuration.
Hence, the fundamental frequency of the torsional motion is obtained from
$\beta = m\omega_0^2/2$, where $m$ is the moment of inertia and
its value is fixed as $m=1$.
In the transition state ($x=0$), the potential is zero.
The torsional energy barrier separating the isomers is calculated as 
$E_b = V_g(0) - V_g(\pm x_0) = m\omega_0^2x_0^2 / 8$. 
In this work the energies are scaled with respect to $\omega_0$, so
we use $\beta = 1/2$.
Calculations at the level of density functional theory estimated
a isomerization barrier of $16.7$ Kcal/mol\cite{Grein} ($5841$ cm$^{-1}$), 
which is much larger than that in the unsubstituted biphenyl molecule
($E_b \sim 500$ cm$^{-1}$\cite{Takei}). 
In the biphenyl substituted with methyl
groups there will be many more states belonging to the reactant and
product, making it an excellent example of the capabilities of the
parallel transfer control mechanism\cite{par1,par2}.

For the fundamental frequency of the torsional motion, $\omega_0$, we
use data from biphenyl, where $\omega_0 = 70$ cm$^{-1}$\cite{Takei}, 
and apply a mass correction due to the different reduced masses
of the rings in the dimethyl biphenyl $\mu_d$ and the biphenyl molecule 
$\mu_0$, $\mu_d / \mu_0 = 1.195$. Hence, we assume for the 2,2'-dimethyl
biphenyl $\omega_0 \approx 64$ cm$^{-1}$, and thus
$E_b / \omega_0 \sim 91.3$ and $x_0 = 27$.

There is relatively few detailed information concerning the excited
electronic states of the dimethyl biphenyl molecule, but the peak of the 
$B$ band ($\lambda_\mathrm{max} \sim 227$ nm\cite{Suzuki}) 
is not very displaced from that in biphenyl ($\lambda_\mathrm{max} \sim 237.5$ 
nm\cite{biphenyl1,biphenyl2,biphenyl3}), 
although the peak intensity is smaller.
In biphenyl, the excited state has a minimum near $\theta = 0$
and the potential is very flat around the equilibrium geometry\cite{biphenyl1,biphenyl2,biphenyl3}.
We model the excited state in the 2,2'-dimethyl biphenyl potential
as a barrierless quartic potential
\begin{equation}
V_e(x)= \alpha x^4 ,
\label{potex}
\end{equation}
with $\alpha = 2\cdot 10^{-4}$.  
The energy gap between the electronic states is included
in the frequency of the pulse.

To construct the Hamiltonian matrix we use the first $100$ localized
eigenstates of the ground potential ($50$ belonging to each isomer)
and $121$ levels in the excited potential (those with quantum numbers
from $v'= 180$ to $v'=300$) and apply the FGH procedure.

\subsection{Geometrical Optimization}

In the ground electronic state, the system has a set of 
(localized) vibrational states that belong to isomer A, $|A,j\rangle$, and 
a set of (localized) vibrational states that belong to isomer B, $|B,k\rangle$.
We are only interested in stable isomers. In addition, there will be
(delocalized) vibrational states with energy larger than the internal
isomerization barrier. These states will not be used in the geometrical
optimization procedure.
The reaction will proceed through the excited electronic
state E with delocalized vibrational states $|E,l\rangle$.  We are
interested in maximizing the overall yield of the 
isomerization reaction, defined as
\begin{equation}
\chi_i = \sum_{k\in B} \left| \langle B, k | {\sf U}(T,0;\epsilon_p,
\epsilon_S) | \psi^A_i \rangle \right|^2
\label{yield}
\end{equation}
where the initial state $|\psi^A_i\rangle = \sum_j^{N_c} a_{ji} |A,j\rangle$ 
is a superposition state involving $N_c$ vibrational levels of the initial 
isomer. 
We sum over all
the vibrational levels that belong to $B$ (that is, the localized
states, abbreviated as $k \in B$) 
at final time, $T$.
The propagator ${\sf U}(T,0;\epsilon_p,\epsilon_S)$ 
depends on two external fields, the pump pulse $\epsilon_p(t)$ and the
dump (or Stokes) pulse $\epsilon_S(t)$. 
For the derivation of the more general equations, we
assume that each field drives a different transition, 
although for the
2,2'-dimethyl-biphenyl or any other symmetrical system,
both pulses couple each isomer to the excited state. Then the propagators
will depend on a field $\epsilon_0(t) = \epsilon_p(t) + \epsilon_S(t)$.
In general we will omit the-time dependence for brevity.
For each superposition state $|\psi^A_i\rangle$ one obtains a
different yield, $\chi_i$.

We want to find $|\psi^A_i\rangle$ that maximizes $\chi_i$
for a given set of pulses $\epsilon_p$, $\epsilon_S$. 
If we are only allowed to change the initial state,
the amplitudes of the superposition can be obtained using a variational
method.
The result gives the matrix equation\cite{par2}
\begin{equation}
\sum_{k\in B}^{N_c} \left( F_{jk} -\chi_i\delta_{jk} \right) a_{ki} = 0
\end{equation}
where $\delta_{jk}$ is the Kronecker delta and the operator ${\sf F}$
has matrix elements
\begin{equation}
\begin{array}{cl}
F_{jk} =  \sum_{n\in N_B} & \langle A,j |{\sf U}^\dagger(T,0;\epsilon_p,
\epsilon_S) | B,n\rangle \\
 & \times \langle B,n | {\sf U}(T,0;\epsilon_p, \epsilon_S) | 
A,k \rangle \end{array}
\end{equation}
However, if the time-evolution is split in different time-intervals,
it is possible to geometrically optimize the wave function in
between. This is most natural when the optimization is performed
over a well defined set of states. For instance, in the isomerization
reaction, if the pump- and Stokes-pulses are time-delayed, it is
natural to ask how the yield of the reaction can be improved if
the wave function in the excited state can be prepared after $\epsilon_p$
and before $\epsilon_S$ acts. 

We first rewrite in detail the sum over the excited states
$|E,l\rangle$ in the yield
$$\chi_{i} = \sum_{k \in B} \left| \sum_{l\in E} \langle \psi^A_i | 
{\sf U}_1^{\dagger} (T_m,0;\epsilon_p) | E, l \rangle \right.$$
\begin{equation}
\left.  \langle E,l | {\sf U}_2^{\dagger}(T,T_m;\epsilon_S)   
 |B,k \rangle  \right|^2 , 
\end{equation}
where $T_m$ is the time at which $\epsilon_p(t)$ is switched off
and the indices $1$ and $2$
refer to the propagators that depend on the first and second pulses,
respectively.
Then we allow ourselves to change the intermediate state before $\epsilon_S$,
by which we start the stimulated emission in the so-called ``bridge state'' 
$|\psi_b^E\rangle$, instead of the state prepared after $\epsilon_p$, 
$|\psi^E(T_m)\rangle$.
We now rename $\chi_{ib}$ to the isomerization yield, as it
depends on two wave functions, 
$|\psi^A_i\rangle$ and $|\psi^E_b\rangle$, and rewrite the equation as
$$\chi_{ib} =  \sum_{l \in E} \left| \langle \psi^A_i |{\sf U}_1^{\dagger}
(T_m,0;\epsilon_p) |E, l\rangle \right|^2 \times$$
\begin{equation}
\sum_{k \in B} \left| \langle \psi^E_b |{\sf U}_1^{\dagger}
(T_m,0;\epsilon_p) |B, k\rangle \right|^2 = \chi_i \chi_b
\end{equation}
which is a product of the yields for each one-photon process.
More restrictive optimization procedures are also possible\cite{Note}.

By the geometrical optimization procedure we can maximize the yield
with respect to $|\psi^A_i\rangle$ and 
$|\psi^E_b\rangle$, the latter being a superpositon of
$N_E$ vibrational levels of the $E$ state,
$|\psi^E_b \rangle = \sum_{l\in N_E} e_{lb} |E,l\rangle$.
It is important to notice that the optimization of
$|\psi^E_b\rangle$ requires some preparation time $T_p$ that is included 
in ${\sf U}_2(T,T_m;\epsilon_S)$. Because we use the GeOp procedure and
therefore we do not treat the process dynamically, $T_p$ is not defined. 
Hence, the time-delay between the pump and Stokes is not defined either.

We obtain $|\psi^E_b\rangle$  by maximizing the one-photon
transition from the $E$ to the $B$ state via ${\sf U}_2(T_m,T;\epsilon_S)$,
yielding the eigenvalue equation
\begin{equation}
\sum_{k\in E}^{N_E} \left( G_{jk} -\chi_b 
\delta_{jk} \right) e_{kb} = 0
\end{equation}
where
\begin{equation}
G_{jk} = \sum_{k\in B}  \langle E, j | {\sf U}^{\dagger}_2(T,T_m;\epsilon_S) 
| B, k\rangle \langle B, k | {\sf U}_2(T,T_m;\epsilon_S) |E, k \rangle
\end{equation}
On the other hand, the initial state $|\psi_i^A\rangle$
is found by maximizing the transition probability 
from $A$ to $E$ conditioned on the choice of $\epsilon_p$. 
It must satisfy the equation
\begin{equation}
\sum_{k\in A}^{N_A} \left( F_{jk} -\chi_i\delta_{jk} \right) a_{ji} = 0
\end{equation}
where
\begin{equation}
F_{jk} = \sum_{k\in N_E}  \langle A, j | {\sf U}^{\dagger}_1(T_m,0;\epsilon_p) 
| E, k\rangle \nonumber \\
\langle E, k | {\sf U}_1(T_m,0;\epsilon_p) |A, i \rangle
\end{equation}
Finally, the yield for the isomerization reaction is obtained as
$\chi_{ib} = \chi_i \chi_b$.
Unless otherwise speficied, we always choose the number of levels 
that participate in each of the optimizations, such
that they span the same energy bandwith, {\it e.g.} if $n_i$ and $n_f$ and
$m_i$ and $m_f$ are the smaller/larger quanta within $N_A$ and $N_E$
respectively, we constrain the levels that we choose in both maximizations
so that their energy differences are equal, $E(n_f) - E(n_i) =
E(m_f) - E(m_i)$.

In the case of 2,2'-dimethyl-biphenyl the previous equations remain valid
by simply changing $\epsilon_p$ and $\epsilon_S$ by $\epsilon_0$ which
is the sum of both.
Now ${\sf U}_1$ and ${\sf U}_2$ are the same propagators.
However,
$G_{jk}$ and $F_{jk}$ are not equal because of the different role
of the initial and final states in Eqs.(7) and (9).
In addition, in this case it is possible to use a single long pulse
$\epsilon_0(t)$ that could drive both the pump and dump processes.

\section{Controlling the dimethyl-biphenyl isomerization}

\subsection{Numerical Results}

We integrate the time-dependent Schr\"odinger equation (TDSE)
in the energy representation for the $1D$-model
Hamiltonian with different parameters of the pulses 
(peak amplitude and duration). 
We use Gaussian pulses and
the pulse frequencies are chosen to resonantly excite an excited state
$|E,k\rangle$ ($k = 220$) that maximizes the Franck-Condon factor with the
ground state level, $|\langle E, k|A,0\rangle|^2$.
The dipole matrix element $\langle E, 220 | A, 0 \rangle = 
0.109$, while $\langle E, 220 |B, 0 \rangle = -0.109$ (the sign
depends on the numerical procedure since the symmetrized states
are degenerate).

The reaction yield (final population in $B$) is represented as a function
of the pulse area\cite{Shore} (omitting the factor $ea_0/\hbar$)
\begin{equation}
{A_0} = \int_{-\infty}^{\infty} \epsilon_0(t) dt
\end{equation}
which parametrizes the integrated strength of the coupling, as discussed
in detail in the following section. Hence, when comparing results obtained
with pulses of different duration at a given $A_0$ one must take into
account that the actual amplitudes differ. Since the energy
of the pulse depends on the intensity and duration, the results
obtained using shorter pulses (at a fixed $A_0$) imply using stronger pulses.

\begin{figure}
\includegraphics[width=6.8cm,scale=1.0,angle=-90]{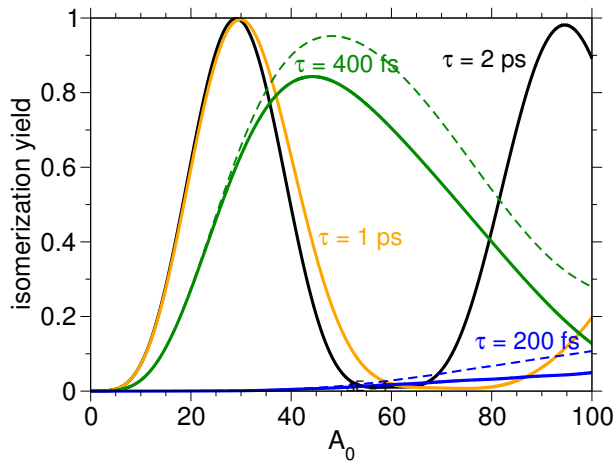}
\caption{Yield of isomerization as a function of the pulse area $A_0$
of the single long pulse that drives the reaction, for different pulse
durations (solid lines). 
Also shown are results obtained when the initial
wave function is a superposition state optimized to maximize the 
transfer (dashed lines). Using long pulses the yield can barely improve since
essentially one one quantum pathway links the same initial and final states,
and therefore there is no interference.}
\end{figure}

First, we study the effect of pulses of duration
(full width half maximum, fwhm) $\tau = 200$ fs, $400$ fs, $1$ and $2$ ps
starting in $|A,0\rangle$.
The isomerization reaction is driven by a single pulse, which is possible since 
both isomers have the same energy.
In Fig. 1 we show the yield of isomerization at 
the end of the pulse starting in the
ground vibrational level.
Since the energy spacing between adjacent levels is 
$64$ cm$^{-1}$ in the ground state (smaller in the excited state),
using $1$ or $2$ ps pulses the states are energy-resolved and the system
behaves mainly as a $3$-level system in so-called $\Lambda$ configuration. 
That is, basically only $3$ levels participate and the dynamics show 
some kind of Rabi oscillations.
As discussed in the following
section [Sec.III.b] the oscillations follow a sine to the fourth 
behavior, which is characteristic of two-photon (or pump-dump) processes.

However, with shorter pulses ($\tau = 400$ fs) one first observes deviations 
at larger intensities that produce asymmetries in the Rabi oscillation. 
For even shorter pulses the population transfer to the isomer $B$ is
basically blocked. Using strong fields ($A_0 \sim 30$) 
one observes first substantial population in the
excited intermediate state ($\sim 70$\%) and then Raman excitation.

It is known that, using longer pulses, the optimization of the
initial state does not improve the results beyond the possible
benefit of exploiting larger transition-dipole matrix-elements,
since there are no possible interfering pathways\cite{par3}. 
We have applied the GeOp procedure to the dynamics driven by the shorter pulses,
optimizing the initial wave function as a superposition of the $2$ or
$4$ lowest vibrational levels for the $\tau = 400$ and $\tau = 200$ fs
pulses, respectively.
The results with even shorter pulses ($\tau \le 40$ fs) show yields
smaller than $0.005$ even considering GeOp with a large number ($10-20$)
of vibrational levels, hence they are not shown in the figure.
Clearly, the isomerization reaction cannot be driven efficiently by
a single ultrashort pulse.

\begin{figure}
\includegraphics[width=7.5cm,scale=1.0]{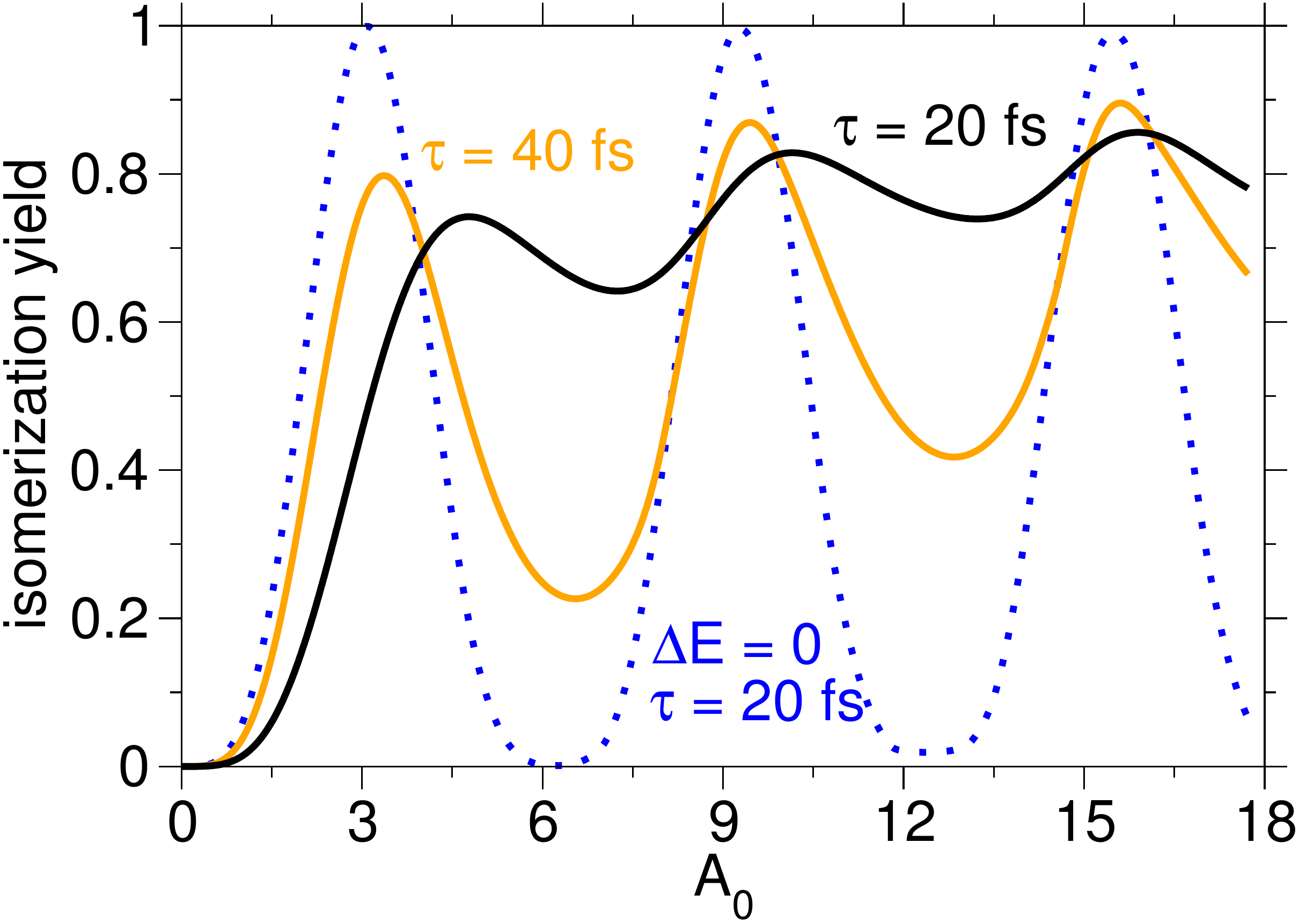}
\caption{Yield of isomerization as a function of the pulse area $A_0$
with pulses of different durations
for a pump and dump process (using two identical time-delayed pulses)
where both the initial state and the intermediate state are optimized
to maximize the yield of each respective optical transition.
The dotted line represents the results where we model the system with
degenerate vibrational levels, as explained in the text.}
\end{figure}

In order to drive the isomerization reaction 
more effectively with shorter pulses
we must split the driving pulse in two, and optimize both the initial 
$|\psi^A_i\rangle$ and the intermediate state $|\psi^E_b\rangle$. 
Since the excited
state is optimized, $|\psi^E_b\rangle$ is not necessarily dynamically connected
to the state prepared after the pump pulse, 
$|\psi^E(T_m)\rangle = U_1(T_m,0;\epsilon_1)|\psi^A_i\rangle$. 

In Fig. 2 we show the results of the optimization using pulses of
$40$ and $20$ fs durations (fwhm) and different pulse amplitudes,
parametrized as a function of the pulse area [Eq.(10)].
For the results with $40$ fs pulses we use the lowest $13$ vibrational
levels in $A$ to optimize $|\psi^A_i\rangle$ and $20$ levels in $E$ around
$v = 220$, such that approximately both sets of levels span the
same bandwidth, which is approximately the pulse bandwidth. 
For the optimization with $20$ fs pulses we use $25$ levels 
in $A$ and $37$ in B. Using the same conditions, we also show the
results that would be obtained using a simplified model Hamiltonian
where we keep the same couplings, but make all states degenerate.
In this model one can obtain insightful analytical results\cite{par3}.

The yield of isomerization shows oscillations. Comparing these 
oscillations with those obtained using a single long pulse, 
we observe that with two pulses
(optimizing the $|\psi^E_b\rangle$ state) one can achieve high yields
at significantly lower pulse areas. This is a feature of 
parallel transfer, where absorption (or stimulated emission) 
can be accelerated by maximizing the use of the effective
transition dipole between the electronic states. The required
area can roughly decrease with the number of levels that participate.
On the other hand, the results are relatively similar for the
$20$ and $40$ fs pulses, although the maximum yields can be slightly
larger (but more sensitive to the area) for the shorter pulse.
This reflects some results obtained in one-photon processes, 
where the parallel transfer depends more crucially on the
initial phases for shorter pulses, imposing some kind of
generalized pulse area theorem\cite{par2}. 
For longer pulses dynamical phases become more important. The
maximum yields are smaller but the sensitivity to the area
decreases. In the opposite extreme, by removing the dynamical
phases (imposing degenerate states in the Hamiltonian)
one reproduces perfect Rabi oscillations, characteristic of a
simple $3$-level system. We explain the process in more
detail in the following section.

\begin{figure}
\includegraphics[width=7.2cm,scale=1.0,angle=-0]{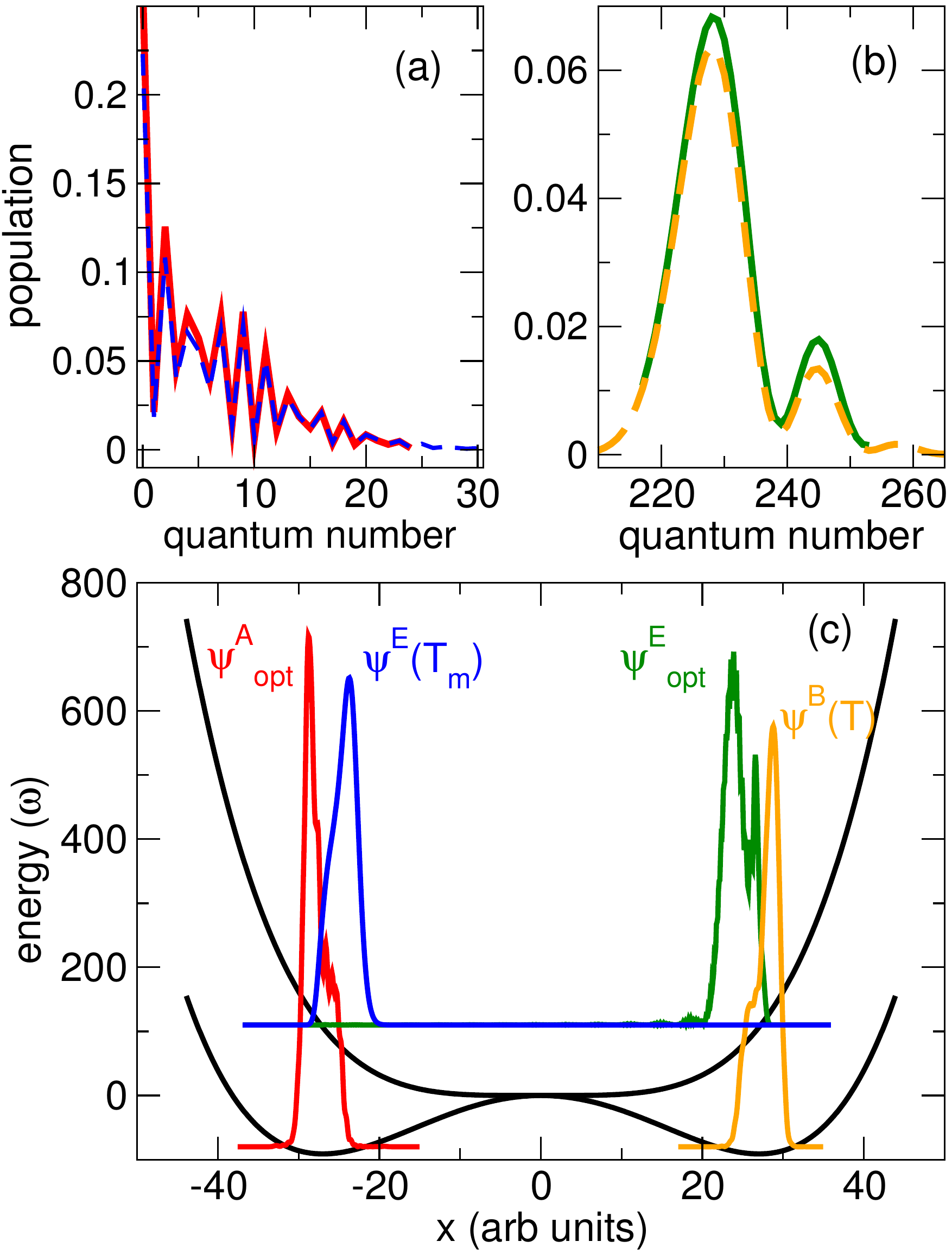}
\caption{In (a) we compare the optimal initial wave packet $|\psi^A_\mathrm{opt}
\rangle$ (orange line) with the final state prepared after the dump pulse
in the B isomer $|\psi^B(T)\rangle$ (black dashed line) 
in the energy representation (vibrational quantum numbers). 
In (b) we compare the optimal bridge state $|\psi^E_\mathrm{opt}\rangle$ (orange line) 
with the wave packet prepared after the pump pulse (black dashed line).
In (c) we show the different wave packets in the position representation.
See the text for more information.}
\end{figure}

To understand the results of the optimization procedure
we need to consider the geometrical mechanism behind the process.
In Fig. 3 we compare the state that is prepared after the first pulse
(at time $T_m$), $|\psi^E(T_m)\rangle$, with the optimized 
intermediate state $|\psi^E_\mathrm{opt}\rangle$. 
Because we limit the number of 
states that can participate in the superposition, the populations in 
$|\psi^E_\mathrm{opt}\rangle$ are rescaled with respect to those in 
$|\psi^E(T_m)\rangle$,
but otherwise are similar. 
The effect of the optimization is basically to change
the relative phases of the wave packet components.
The effect can be visualized in the coordinate representation
as a displacement of the wave packet from near the Franck-Condon region
of the $E \leftarrow A$ transition, to near the Franck-Condon region of the 
$B \leftarrow E$ transition. The optimization in the intermediate state
is essentially equivalent to adding a time-delay that allows the wave packet 
prepared at time $T_m$ to move from one region of the potential to another,
plus additional corrections that minimize the wave packet 
spreading that would occur in free evolution. 

The same analysis can be performed on $|\psi^A_\mathrm{opt}\rangle$ and the 
final wave function that is prepared in the $B$ isomer, $|\psi^B(T)\rangle$.
Because the system is very symmetrical, both states are also very similar,
showing the validity of the analytical solutions and the underlying
geometrical features that govern the GeOp procedure.

In Fig.3(c) we reconstruct the wave functions in the generalized coordinate
$x$, $|\langle x | \psi \rangle|^2$, from the amplitudes and the eigenstates
obtained using the FGH method. As indicated, the optimization amounts
to a displacement of the excited wave function.
Although the parallel 
transfer increases the rate of absorption from $A$ to $E$ and stimulated
emission from $E$ to $B$, because the limiting factor is given by the
free evolution in the intermediate state (or alternatively the preparation
of the optimal intermediate state) there is little gain in the overall rate 
of the isomerization.

\subsection{Simple analytical models}

In the long-pulse, weak-interaction regime, as in the results of Fig. 1
(particularly for $\tau = 1$ ps) the $1D$ molecular model can be
approximated by a simple $3$-level system, where the initial and
final states are both coupled by the same field $\epsilon_0(t)$ with the 
intermediate excited state. Under these conditions the TDSE equation can be 
integrated analytically in the rotating wave approximation (RWA),
where only the {\it in-phase} component of the periodic oscillation of 
the laser $\cos (\omega t) \approx \exp(\pm i \omega t)/2$, is kept (the 
negative part in the absorption and the positive in the stimulated emission).
The final population in $|B,0\rangle$
(which is equal to the yield of isomerization in this model) is
\begin{equation}
\chi = |\langle B,0 | \psi(\infty) \rangle |^2 = \sin^4\left( 
\mu {A_0}/ \sqrt{8} \right)
\end{equation}
where $\mu$ is the absolute value of the transition dipole between the
states, $\mu \sim 0.109$.
This result implies a minimum pulse area of 
$A_0 = \sqrt{2} \pi / \mu$ in
order to reach the maximum yield of isomerization. 

For comparison, if both transitions were independent, the reaction
could proceed sequentially, first from $|A,0\rangle$ to the intermediate
state and then from the intermediate state to $|B,0\rangle$. The
simple two level system implies population transfer that depends as
$\sin^2(A_0 / 2)$ (the Rabi formula) given a minimum pulse area of $\pi$ for
each transition, and a total area of $2\pi$ for the two pulses
to drive the isomerization reaction.

To understand the results using shorter pulses we must address several
questions. The first one is why the isomerization reaction cannot be driven 
by a single (ultrashort) pulse. The second one is how the absorption and
stimulated emission processes can be accelerated, that is,
how the Rabi oscillations of the yield depend on the number of levels
in the superposition.
In the context of the model in the energy-representation, the first
question can be seen to depend on some features of the signs of the
Franck-Condon factors, some of which are related to the symmetry
of the system. The second is a general feature of parallel transfer.

For the optical isomer one can find fundamental symmetry rules
based on the symmetrized eigenfunctions. 
The excited states, $|E,j\rangle$, are symmetric or antisymmetric
with respect to parity for even and odd values of $j$, respectively.
For the ground electronic potential, one can create states with
well defined symmetry:
$|+,j\rangle = \left( |A,j \rangle + |B,j\rangle \right)/\sqrt{2}$
and $|-,j\rangle = \left( |A,j \rangle - |B,j\rangle \right)/\sqrt{2}$.

To uniquely correlate the sign with the parity for all possible values
of $j$, the wave function in $B$ must flip its sign for even $j$. 
Because the signs of wave functions are arbitrary this global gauge fixing 
is allowed. 
The matrix elements are different depending on the choice
of phases, but the final results of the dynamics do not change.
Here we adopt this choice to simplify the analysis.
Then some general properties of the  matrix elements can be easily calculated. 
For symmetric transition dipoles (as in the Condon limit),
$\langle E, k | \mu | +, j \rangle = [1+(-1)^k]/2$ and
$\langle E, k | \mu | -, j \rangle = [-1+(-1)^k]/2$. Writing
$|A,j\rangle$ and $|B,j\rangle$ in terms of $|+,j\rangle$ and
$|-,j\rangle$,
$|A,j\rangle = \left(|+,j\rangle + |-,j\rangle\right)/\sqrt{2}$,
$|B,j\rangle = \left(|+,j\rangle - |-,j\rangle\right)/\sqrt{2}$,
one can obtain the fundamental symmetry rule for the 
nondiagonal Hamiltonian elements in terms of the localized states,
\begin{equation}
\langle E, k | \mu | A, j\rangle = (-1)^k \langle E, k | \mu | B, j\rangle
\label{symrule}
\end{equation}
which implies that the sign of the matrix elements only depends
on the parity of the $|E,k\rangle$ states. 
For antisymmetric transition dipoles one must multiply the
second term by $(-1)^{k+1}$.

Now assume, for simplicity, that the dipole matrix elements
have equal magnitude and only differ in sign, which 
only depends on the vibrational level $|E,k\rangle$,
$\langle E, k | \mu | A, i\rangle = \mathrm{sign}_1(k) \mu$,
$\langle E, k | \mu | B, j\rangle = \mathrm{sign}_2(k) \mu$
(alternatively consider first that there is only one active
vibrational level in $A$ and $B$) and that all levels are
degenerate, as in the model tested in Sec.3A.
Then starting in $|A,0\rangle$, the pulse prepares
a superposition state $|\psi^E(T_m)\rangle = \sum_k \mathrm{sign}_1(k)
|E,k\rangle / \sqrt{N_E}$ (where $N_E$ is the number of accessible
levels in the excited state). The amplitude in this superposition
will depend on the pulse parameters, but all the population
in the excited state will be in $|\psi^E\rangle$, as 
$$\langle \psi^E(T_m) | \mu | A, i \rangle = \mu \sum_k \mathrm{sign}_1(k)^2 
/ \sqrt{N_E} = \sqrt{N_E} \mu $$ 
is the only non zero matrix element of the Hamiltonian, 
or alternatively, $|\psi^E\rangle$ is the only bright state.

In addition, in the absence of dynamical phases, the superposition
does not change. Its coupling with any vibrational level in the
$B$ isomer is
$\langle B, j | \mu | \psi^E(T_m) \rangle = \mu^2
\sum_k \mathrm{sign}_2(k) \mathrm{sign}_1(k) = \mu^2
\sum_k (-1)^k = 0$.
Hence a single ultrafast pulse cannot move the population directly from
the $A$ isomer to the $B$. This will occur whenever there are at least two
$k$ levels involved in the sum (it does not restrict population transfer
using long pulses). The dynamical phases, however, can change the
sign of the superposition such that it is no longer orthogonal to
the $|B,j\rangle$ levels. But this change depends on the mass of the system
and induces, in the coordinate representation, the motion of the wave packet
$|\psi^E(t)\rangle$.

On the other hand, in order to accelerate the transition from $A$ to 
$E$ and to avoid the Raman decoupling (Autler-Townes splitting) induced
by the initially unpopulated vibrational levels in $A$, one just
needs to prepare an initial superposition state in the form
$|\psi^A_i\rangle = \sum_j |A, j\rangle / \sqrt{N_A}$,
where $N_A$ is the number of accessible levels in $A$.
This state will also prepare $|\psi^E(T_m)\rangle$ but the coupling
$$\langle \psi^E(T_m) | \mu | \psi^A_i \rangle = \mu
\sum_j \sum_k \mathrm{sign}_1(k)^2 / \sqrt{N_A N_E} =$$ 
$=  \sqrt{N_A N_E} \mu$,
increases with the number of participating levels in the superposition.
The same argument applies to optimize $|\psi^E_b\rangle$ in order to accelerate
the transition to $B$. However, as discussed, $\langle \psi^E_b |
\psi^E(T_m) \rangle = 0$.

Although for enantiomers this orthogonality is enforced by symmetry, 
reflected in the signs of the dipole matrix elements,
the negligible overlap of the excited state prepared in the Franck-Condon
region of the $A$ isomer with respect to the $B$ states is a rather general
geometrical principle that stems from the fact that both equilibrium 
configurations
are spatially separated. As the results in Fig. 3 show, the geometrical
optimization can find the superposition states that accelerate the
absorption 
from isomer $A$ to the excited state and the stimulated
emission from $E$ to isomer $B$. These wave functions sit in their
respective Franck-Condon regions and do not overlap. Hence it is not
possible to accelerate the whole isomerization reaction.
The optimization of $|\psi^E_b\rangle$ on the other hand, basically amounts
to waiting for the wave packet prepared in $E$ after the first pulse ends
$|\psi^E(T_m)\rangle$ to reach, by free evolution in $B$, to the other 
Franck-Condon window. This process is limited by the dynamical phases
(mass of the system) and cannot be accelerated.
In short, what the GeOp scheme discovers is the well-known pump-dump process.

\section{Conclusions}

We have studied the laser-driven isomerization reaction of an 
oriented 2,2'-dimethyl biphenyl molecule through an excited barrierless
electronic state using the recently developed Geometrical Optimization
procedure. Our goal was to analyze
whether an initial wave packet in the ground state, with optimized
amplitudes and phases, could be used to enhance the yield of the reaction 
at faster rates, exploring how the geometrical restrictions induced by the 
symmetry of the system impose limitations in the optimization procedure.
We used a very simple general $1$-D model for the system, based on symmetric
quartic potentials.
The study omits further interesting considerations involving the neglected
dimensions of the system, and hence, possible competing processes such
as IVR or intersystem crossing at conical intersections, which will likely
play a role in a more realistic model of the dynamics, even using short
pulses.

We have found that using long (picosecond) pulses the reaction can be         
driven by a single pulse and the results are not sensitive to the initial state 
coherences. Hence, the geometrical optimization procedure is not necessary
in this limit as no PT is possible.

On the other hand, using shorter (femtosecond) pulses, the reaction must be
driven by a pair of pulses. Since the system is symmetric, we use a pair
of identical pulses, which operate as the pump- and dump-pulses. 
This leads to optimizing both the initial wave function,
as well as the wave packet prepared in the excited state, which we
called the bridge state.
The results showed that the optimal wave functions 
are approximately copies of the initial wave functions, 
displaced to the Franck-Condon regions of their respective
pump- and dump-processes. Essentially, the geometrical optimization of the 
bridge state represents (and hides) the dynamical process of the free
evolution of the wave packet in the excited electronic state.
Hence the GeOp procedure can only optimize and accelerate the electronic 
absorption and the stimulated emission.
However, the isomerization rate is primarily determined by the
torsion of the rings (or the equivalent geometrical process for the
isomerization reaction under study), which 
does not proceed through PT and remains the main limitation of the dynamics.

\section*{Acknowledgment}
This work was supported by the Korean government through 
the Basic Science Research program (NRF-2013R1A1A2061898) and
the EDISON project (2012M3C1A6035358),
by the Spanish government through the MICINN projects 
CTQ2012-36184 and CTQ2015-65033-P and by the COST XLIC Action CM1204.


\begin{thebibliography}{99}
\bibitem{Note}
An interesting different procedure requires finding the variational
optimization of the yield
\begin{equation}
\chi_{ib} = 
\sum_{k \in B} \left| {\langle \psi^A_i | {\sf U}_1^{-1}(T_m,0;\epsilon_p) 
| \psi^E(T_m) \rangle \langle \psi^E_{b} | {\sf U}_2^{-1}(T,T_m;\epsilon_S)   
 |B,k \rangle } \right|^2
\end{equation}
with the constraint that  $|\psi^E_b\rangle$ and $|\psi^E(T_m)\rangle$ 
are related by a diagonal unitary matrix, such that the state
that initiates the stimulated emission from E differs only from
the state prepared after the pump pulse by relative (dynamical) phases,
not by the vibrational populations.
The maximum yields are always smaller (or at most equal) than those
obtained using the simpler method of Eq.(7). However, the physical
process that allows to implement the control is simpler. 
The optimization of $|\psi^E_b\rangle$ is equivalent to finding the
optimal time delay between the pulses.
\end{thebibliography}
\end{document}